\documentclass[runningheads]{llncs}
\usepackage{graphicx}
\usepackage{amsmath}
\usepackage{amsfonts}
\usepackage{amssymb}
\usepackage{hyperref}
\usepackage{proof}
\usepackage{xspace}
\usepackage{caption}
\usepackage{stmaryrd}
\usepackage{framed}
\usepackage[dvipsnames]{xcolor}
\usepackage{colortbl}
\usepackage{mdframed}
\usepackage[linesnumbered,ruled,vlined]{algorithm2e}
\DontPrintSemicolon

\newcommand{\authorA}{Dejan Jovanovi\'{c}\xspace}
\newcommand{\authorB}{Bruno Dutertre\xspace}
\newcommand{\institution}{SRI International\xspace}

\newcommand{\kind}{\textsc{kind}\xspace}
\newcommand{\pdkind}{\textsc{pdkind}\xspace}
\newcommand{\icccnra}{\textsc{ic3-nra}\xspace}
\newcommand{\yices}{\textsc{yices2}\xspace}
\newcommand{\libpoly}{\textsc{libpoly}\xspace}
\newcommand{\sally}{\textsc{sally}\xspace}
\newcommand{\mcsat}{\textsc{mcsat}\xspace}
\newcommand{\isat}{\textsc{isat3}\xspace}
\newcommand{\nuxmv}{\textsc{nuxmv}\xspace}

\DeclareMathOperator{\proot}{\mathsf{root}}

\DeclareMathOperator{\sgn}{\mathsf{sgn}}
\DeclareMathOperator{\sgncstr}{\mathsf{sgncstr}}

\DeclareMathOperator{\btrue}{\top}
\DeclareMathOperator{\bfalse}{\bot}

\DeclareMathOperator{\system}{\mathfrak{S}}
\newcommand{\init}{\ensuremath{I}}
\newcommand{\trans}{\ensuremath{T}}

\SetKw{Unsat}{unsat}
\SetKw{Sat}{sat}
\SetKw{True}{true}
\SetKw{False}{false}

\SetKwFunction{SolverCheck}{check}
\SetKwFunction{SolverPropagate}{unitPropagate}
\SetKwFunction{SolverAnalyze}{analyzeConflict}
\SetKwFunction{SolverAnalyzeFinal}{analyzeFinal}
\SetKwFunction{SolverDecide}{decideValue}
\SetKwFunction{SolverInConflict}{inConflict}
\SetKwFunction{SolverBacktrackWith}{backtrackWith}
\SetKwFunction{SolverAssert}{assert}
\SetKwFunction{OwnerOf}{ownerOf}
\SetKwFunction{TrailLevel}{level}
\SetKwFunction{TrailSize}{size}

\newcommand{\ISolverCheck}[1]{\textsc{solver::check(#1)}}
\newcommand{\ISolverAssert}[1]{\textsc{solver::assert(#1)}}

\SetKwFunction{VarQueuePop}{pop}
\SetKwFunction{VarQueueEmpty}{empty}
\SetKwFunction{VarQueueInsert}{insert}

\SetKwFunction{TrailLevel}{level}
\SetKwFunction{TrailSize}{size}

\newcommand{\trail}[1]{\llbracket\; #1 \;\rrbracket}
\newcommand{\decision}[2]{\ensuremath{{#1} \mapsto #2}}
\newcommand{\propagation}[3]{\ensuremath{#1 \overset{#3}{\leadsto} #2}}

\newcommand{\evaluates}[3]{\ensuremath{\mathsf{evaluates}[#1](#2, #3)}}

\newcommand{\cadcell}[2]{\ensuremath{\mathsf{describeCell}(#1, #2)}}
\newcommand{\cadcellbasic}[2]{\ensuremath{\mathsf{describeCellBasic}(#1, #2)}}

\newcommand{\emphcolor}{BrickRed}

\newmdtheoremenv{extension}{Modification}

\begin{document}
\title{Interpolation and Model Checking for Nonlinear Arithmetic%
\thanks{
This material is based upon work supported by the Defense Advanced Research
Project Agency (DARPA) and Space and Naval Warfare Systems Center, Pacific (SSC
Pacific) under Contract No. N66001-18-C-4011, and the National Science
Foundation (NSF) grant 1816936. Any opinions, findings and conclusions or
recommendations expressed in this material are those of the author(s) and do not
necessarily reflect the views of DARPA, SSC Pacific, or the NSF.
}}
\author{\authorA \and \authorB}
\authorrunning{\authorA et al.}
\institute{\institution}
\maketitle              %
\begin{abstract}
We present a new model-based interpolation procedure for
satisfiability modulo theories (SMT). The procedure uses a new mode of
interaction with the SMT solver that we call {\em solving modulo a
  model\/}. This either extends a given partial model into a full
model for a set of assertions or returns an explanation (a model
interpolant) when no solution exists.  This mode of interaction fits
well into the model-constructing satisfiability (MCSAT) framework of
SMT. We use it to develop an interpolation procedure for any
MCSAT-supported theory. In particular, this method leads to an
effective interpolation procedure for nonlinear real arithmetic. We
evaluate the new procedure by integrating it into a model checker and
comparing it with state-of-art model-checking tools for nonlinear
arithmetic.
\end{abstract}

\section{Introduction}
\label{sec:introduction}

Craig interpolation is one of the central reasoning tools in modern
verification algorithms. Verification techniques such as model
checking rely on Craig
interpolation~\cite{craig1957three,mcmillan2014interpolation} as a
symbolic learning oracle that drives abstraction refinement and
invariant inference.  Interpolation has been studied for many
fragments of first-order logic that are useful in practice, such as
linear arithmetic~\cite{henzinger2004abstractions}, uninterpreted
functions~\cite{mcmillan2005interpolating,cimatti2008efficient},
arrays~\cite{mcmillan2008quantified,hoenicke2018efficient}, and
sets~\cite{kapur2006interpolation}. In these fragments, a typical
interpolation procedure constructs interpolants by traversing the
clausal proof of unsatisfiability provided by an SMT
solver~\cite{huang1995constructing,krajivcek1997interpolation,pudlak1997lower}
while performing interpolation locally at proof nodes. A major missing
piece in the class of fragments supported by interpolating SMT solvers
is nonlinear arithmetic,\footnote{By nonlinear arithmetic we mean
Boolean combination of arithmetic constraints over arbitrary-degree polynomials.}  as the
complex reasoning required for nonlinear arithmetic makes fine-grained
symbolic proof generation extremely difficult.

We present an approach to interpolation that is driven by models rather than
proofs. Given a pair of formulas $A$ and $B$ such that $A \wedge B$ is
unsatisfiable, an interpolant is a formula $I$ that is implied by $A$ and
inconsistent with $B$. Recent model-based decision procedures, specifically the
ones developed within the MCSAT~\cite{de2013model,jovanovic2013design} framework
for SMT, are internally naturally interpolating. But, rather than interpolating
two formulas, they provide a way to interpolate a set of constraints against a
partial model.  We capitalize on this internal ability, and extend it so that a
formula $A$ can be checked and interpolated against a partial model (\emph{model
interpolation}).  This is closely related to the ability of modern SAT solvers
to perform solving modulo assumptions~\cite{een2003temporal}, a technique that
can also been used to provide interpolation capabilities in finite-state model
checking~\cite{bayless2013efficient}.

We take advantage of model interpolation to build a
formula-interpolation procedure through a simple idea: we can compute
an interpolant of formulas $A$ and $B$ by iteratively interpolating
(and refuting) all models of $B$ with model interpolants from $A$. We
develop the interpolation procedure within the MCSAT framework.  This
immediately allows us to generate interpolants for any theory
supported by the framework. As MCSAT provides efficient complete
solvers for nonlinear real
arithmetic~\cite{jovanovic2012solving,jovanovic2017solving}, we
develop the first complete interpolation procedure for real nonlinear
arithmetic.

To show that this new interpolation procedure is an effective tool
that can be used on real-world problems, we integrate it into a model
checker that uses interpolation for inferring $k$-inductive
invariants. We evaluate this model checker on a set of industrial
benchmarks. Our evaluation shows that the new procedure is highly
effective, both in in terms of speed, and the ability to support the model checker in its
quest for counter-examples and invariants.

\paragraph{Outline}

Section~\ref{sec::background} gives background on SMT,
interpolation, and nonlinear arithmetic. Section~\ref{sec::solving}
presents solving modulo a model and model interpolation, and develops
the general interpolation procedure. In Section~\ref{sec::nonlinear},
we discuss the particular needs of nonlinear arithmetic. In
Section~\ref{sec::evaluation} we evaluate our implementation on
nonlinear model-checking problems. We conclude in
Section~\ref{sec::conclusion} and provide future research directions.

\section{Background}
\label{sec::background}

We assume that the reader is familiar with the usual notions and terminology of
first-order logic and model theory (for an introduction see, e.g.,
\cite{barrett2018satisfiability}).

\paragraph{Nonlinear arithmetic.}

As usual, we denote the ring of integers with $\mathbb{Z}$ and the field of real
numbers with $\mathbb{R}$. Given a vector of variables $\vec{x}$ we denote the
set of polynomials with integer coefficients and variables $\vec{x}$ as
$\mathbb{Z}[\vec{x}]$. A polynomial $f \in \mathbb{Z}[\vec{y}, x]$ is of the
form \begin{align*} f(\vec{y}, x) = a_m \cdot x^{d_m} + a_{m-1} \cdot
x^{d_{m-1}} + \cdots + a_1 \cdot x^{d_1} + a_0 \enspace, \end{align*} where $0 <
d_1 < \cdots < d_m$, and the coefficients $a_i$ are polynomials in
$\mathbb{Z}[\vec{y}]$ with $a_m \neq 0$.
We call $x$ the \emph{top variable} and the highest power $d_m$ is the
\emph{degree} of the polynomial $f$. As usual, we denote with $f^{(k)}$ the
$k$-th derivative of $f$ in its top variable.
A number $\alpha \in \mathbb{R}$ is a \emph{root of the polynomial} $f \in
\mathbb{Z}[x]$ if $f(\alpha) = 0$.

A \emph{polynomial constraint} $C$ is a constraint of the form
$f\,\triangledown\,0$ where $f$ is a polynomial and
$\triangledown \in \lbrace <, \leq, =, \geq, > \rbrace$. If the
polynomial $f=f(x)$ is univariate then we also say that $C$ is
univariate.
An atom is either a polynomial constraint or a Boolean variable, and
formulas are defined inductively with the usual Boolean connectives
($\land$, $\lor$, $\lnot$). The symbols $\btrue$ and $\bfalse$ denote
true and false, respectively.
In addition to the basic polynomial constraints, we will also be working with
extended polynomial constraints. An \emph{extended polynomial constraint} $F$ is
of the form $x \; \triangledown_r \; \proot(f, k, x)$  where $f \in
\mathbb{Z}[\vec{y}, x]$ and $\triangledown_r \in \lbrace <_r, \leq_r, =_r,
\geq_r, >_r \rbrace$. The semantics of this predicate is the following:
Given an assignment that gives real values $\vec{v}$ to the variables $\vec{y}$, then
the roots of $f(\vec{a}, x)$ can be ordered over $\mathbb{R}$. If the
polynomial $f(\vec{a}, x)$ has at least $k$ real roots and $\alpha_k$ is the
$k$-th smallest root\footnote{For example, $x^2-2$ has two roots. The first root
$-\sqrt{2}$ is the smallest of the two and the second root is
$\sqrt{2}$.} then the constraint is equivalent to
$x \; \triangledown \; \alpha_k$.  Otherwise, the constraint evaluates
to $\bfalse$. For example, the constraint $x < \proot(x^2 - 2, 2, x)$
represents $x < \sqrt{2}$.

Given a formula $F(\vec{x})$ we say that a type-consistent variable assignment
$M = \lbrace \vec{x} \mapsto \vec{a} \rbrace$ satisfies $F$ if the formula $F$
evaluates to $\btrue$ in the standard semantics of Booleans and reals. We call
$M$ a model of $F$ and denote this with $M \vDash F$. If there is such a
variable assignment, we say that $F$ is \emph{satisfiable}, otherwise it is
\emph{unsatisfiable}. If two models $M_1$ and $M_2$ agree on the values of their
common variables, we denote the model that combines $M_1$ and $M_2$ with $M_1
\cup M_2$.

\begin{definition}[Craig interpolant]
Given two formulas $A(\vec{x}, \vec{y})$ and $B(\vec{y}, \vec{z})$
such that $A \wedge B$ is unsatisfiable, a \emph{Craig interpolant} is
a formula $I(\vec{y})$ such that $A \Rightarrow I$ and $I \Rightarrow
\neg B$. We call the pair $(A, B)$ an \emph{interpolation problem}.
\end{definition}

\paragraph{Model checking.}

A \emph{state-transition system} is a pair $\system = \langle \init, \trans
\rangle$, where $\init(\vec{x})$ is a state formula describing the initial
states and $\trans(\vec{x}, \vec{x}')$ is a state-transition formula describing
the system's evolution. Given a state formula $P$ (\emph{the property}), we want
to determine whether all reachable states of $\system$ satisfy $P$. If this is
the case, $P$ is an \emph{invariant} of $\system$. If $P$ is not invariant,
there is a concrete trace of the system, called a \emph{counter-example}, that
reaches $\neg P$.

The direct way to prove that a property $P$ is an invariant of
$\system$ is to show that it is inductive. This requires showing that
$P$ holds in the initial states: $\init \Rightarrow P$, and that it is
preserved by transitions:
$P(\vec{x}) \wedge \trans(\vec{x}, \vec{x}') \Rightarrow
P(\vec{x}')$. As most invariants are not inductive, a key problem in
model checking is to find am \emph{inductive strengthening} of $P$,
that is, a property $P'$ such that $P' \Rightarrow P$ and $P'$ is
inductive.

\begin{example}[Cauchy–Schwarz inequality]
\label{ex::cauchy}
We can frame the Cauchy–Schwarz inequality as a model-checking problem in nonlinear
arithmetic. The inequality is the following
\begin{equation}
\label{eq::cauchy}
  ( \sum_{i=1}^n x_i y_i )^2 \leq ( \sum_{i=1}^n x_i^2 ) ( \sum_{i=1}^n y_i^2 )\enspace.
\end{equation}
As shown in~\cite{gerhold2005procedure}, many inequalities that
involve a discrete parameter (such as $n$ above) can be converted to
model-checking problems. For inequality \eqref{eq::cauchy}, we
construct the transition system $\system_{cs}
= \langle \init, \trans \rangle$ where
\begin{align*}
\init &\equiv (S_1 = 0) \wedge (S_2 = 0) \wedge (S_3 = 0)\enspace,\\
\trans &\equiv (S_1' = S_1 + x y) \wedge (S_2' = S_2 + x^2) \wedge (S_3' = S_3 + y^2)\enspace.
\end{align*}
The variables $S_1, S_2, S_3$ correspond to the sums in
\eqref{eq::cauchy} in order. The two variables $x$ and $y$ of $\system_{cs}$ model 
the variables $x_i$ and $y_i$ from \eqref{eq::cauchy} in 
each iteration of $\system_{cs}$. Proving the inequality amounts to showing that property 
$P_{cs} \equiv (S_1^2 \leq S_2 S_3)$ is an invariant of
$\system_{cs}$.  Property $P_{cs}$ is not inductive on its own, but
property $P_{cs}' \equiv P_{cs} \wedge (S_2 \geq 0) \wedge
(S_3 \geq 0)$ is an inductive strengthening of $P_{cs}$.
\end{example}

Many modern model-checking techniques, specifically those based on SMT
solving, use interpolation as a tool to automatically infer inductive invariants.
In this context, an interpolant can be used to over-approximate a
transition in the context of a spurious counter-example. In addition
to interpolation, the recent class of techniques broadly
termed \emph{property-directed reachability} (PDR) (e.g.,
\cite{hoder2012generalized,komuravelli2016smt,jovanovic2016property}), relies
 on {\em model generalization\/}, which converts a concrete
counter-example state into a set of counter-examples.

\begin{definition}[Generalization]
\label{def::gen}
Given a formula $F(\vec{x}, \vec{y})$ such that $F$ is true in a model
$M$, we call a formula $G(\vec{x})$ a \emph{generalization of $M$} if
$G(\vec{x})$ is true in $M$ and
$G(\vec{x}) \Rightarrow \exists
\vec{y}\;.\;F(\vec{x}, \vec{y})$.
\end{definition}

A PDR model-checking procedure for nonlinear arithmetic requires both
an interpolation and a generalization procedure.

\section{SMT Modulo Models and Interpolation}
\label{sec::solving}

SMT solvers typically provide an API to assert formulas and to check
the satisfiability of asserted formulas. We denote with
\ISolverAssert{$F$} the solver method that adds the formula $F$ to the set of
assertions to be checked by the solver. We denote with \ISolverCheck{} the
solver method for checking satisfiability, with the following contract.

\begin{framed}
\noindent
\ISolverCheck{}: Check satisfiability of asserted formulas
$A$ and
\begin{enumerate}
\item if there is a model $M$ such that $M \vDash A$, return $\langle \Sat, M \rangle$;
\item otherwise return $\langle \Unsat, \emptyset \rangle$.
\end{enumerate}
\vspace{-10pt}
\end{framed}
In this contract, the solver does not return any form of inconsistency
certificate when the assertions are unsatisfiable.\footnote{Some
solvers support proof generation. While proofs are fundamentally
important, we are interested in certificates that can always be
computed and are useful in supporting further analysis. For example,
proof generation for nonlinear arithmetic is still a hard open
problem.} We generalize the standard SMT satisfiability checking
to \emph{SMT modulo models} as follows.

\begin{framed}
\noindent
\ISolverCheck{$M_0$}: Check satisfiability of asserted formulas $A$ and
\begin{enumerate}
\item if there is a model $M \supseteq M_0$ such that $M \vDash A$, return $\langle \Sat, M, \btrue \rangle$;
\item otherwise return $\langle \Unsat, \emptyset, I \rangle$ where $A \Rightarrow I$ and $M_0 \vDash \neg I$.
\end{enumerate}
\vspace{-10pt}
\end{framed}

SMT modulo models allows one to check that a formula is satisfiable
modulo a partial model $M_0$, by seeking a solution that extends
$M_0$. If there is no such solution, the formula $I$ returned as the
certificate of unsatisfiability is a
\emph{model interpolant}: it is implied by the assertions and inconsistent with
$M_0$ (i.e., $I$ evaluates to $\bfalse$ in the model $M_0$). If we
restrict ourselves to Boolean formulas, SMT modulo models reduces
exactly to solving modulo assumptions~\cite{een2003temporal} used in
the SAT community. Although this idea is not completely new, it is the
first time that it is used for interpolation in SMT, as far as we
know.

\subsection{Interpolation}

Before diving into an approach that can support the above mode of satisfiability
checking, we first show how model interpolation can be used to devise a general
interpolation method.

\begin{algorithm}
\caption{\sc interpolate($A$, $B$)}
\label{alg::interpolate}
$S_A$.\SolverAssert{A} \;
$S_B$.\SolverAssert{B} \;
$I$ $\gets$ $\btrue$ \;
\While {\True} {
  $\langle r_B, M_B \rangle$ $\gets$ $S_B$.\SolverCheck{} \;
  \uIf {$r_B = \Unsat$} {
  	\Return $\langle \Unsat, I \rangle$
	}
	$\langle r_A, M_A, I_A \rangle$ $\gets$ $S_A$.\SolverCheck{$M_B$} \;
	\uIf {$r_A = \Sat$} {
		\Return $\langle \Sat, M_A \cup M_B \rangle$
	}
	$I$ $\gets$ $I \wedge I_A$ \;
	$S_B$.\SolverAssert{$I_A$}
}
\end{algorithm}

Algorithm~\ref{alg::interpolate} shows the pseudocode of a procedure
that checks satisfiability and interpolates two formulas $A$ and
$B$. The basic idea is simple: we enumerate models $M_k$ of the
formula $B$, and refute each model $M_k$ with a model interpolant
$I_k$ from $A$. If the process converges and returns \Unsat, we
collect the model interpolants and construct the final interpolant $I
= \bigwedge I_k$. Each interpolant $I_k$ is implied by $A$ because it
is a model interpolant, so $A \Rightarrow I$. Each model of $B$
is refuted by some model interpolant $I_k$, and so $I \Rightarrow \neg
B$. On the other hand, if the process returns $\Sat$, the procedure
has found a common model for $A$ and $B$. The procedure above is
model-driven and modular, in that it checks the formulas $A$ and $B$
independently while only communicating models (from $B$ to $A$) and
model interpolants (from $A$ to $B$).

\begin{lemma}[Correctness]
\label{lm::correctness}
If \textsc{interpolate($A$, $B$)} returns $\langle \Unsat, I \rangle$ then $A
\wedge B$ is unsatisfiable and $I$ is an interpolant for $(A, B)$. If
\textsc{interpolate($A$, $B$)} returns $\langle \Sat, M \rangle$ then $A \wedge
B$ is satisfiable and $M$ is a model of both $A$ and $B$.
\end{lemma}
Note that Lemma~\ref{lm::correctness} does not claim termination of
the procedure. Termination depends on the ability of model
interpolation to produce a finite number of model interpolants that
can eliminate a potentially infinite number of models.

A naive approach to check a formula $A(\vec{x}, \vec{y})$ for satisfiability
modulo a model $M_0 = \lbrace \vec{y} \mapsto \vec{v} \rbrace$ is to use an
interpolating SMT solver. First, encode the model into a formula $F_M \equiv
\bigwedge (y_i = v_i)$. If the formula $A \wedge F_M$ is satisfiable in a model
$M$, so is $A$ and $M \supseteq M_0$. Otherwise, we compute the interpolant $I$
of $A$ and $F_M$.
This naive approach satisfies the requirements
of \ISolverCheck{$M_0$}, but it is limited for the following reasons.
First, theories such as nonlinear arithmetic have complex models and the formula
$F_M$ can be hard to express. As an example, $x \mapsto \sqrt{2}$ can only be
expressed by extending the constraint language to support algebraic numbers, or
by using additional assertions such as $(x^2 = 2) \wedge (x > 0)$.
More important, traditional interpolation provides no guarantees in terms of
convergence of a sequence of interpolation problems. For example, as already
noted in \cite{schindler2018selfless}, $\neg F_M$ would be a valid interpolant
for $A$ and $F_M$. But such an interpolant only eliminates a single model and
could, in general, lead to nontermination of \textsc{interpolate($A$, $B$)}. To
tackle this issue, we require that the procedure \ISolverCheck{} 
produces interpolants general enough to disallow such infinite sequences of model
interpolants. We do this by adopting the convergence approach and terminology of
\cite{schindler2018selfless} to model interpolation as follows.

\begin{definition}[Model Interpolation Sequence]
\label{def::seq}
Given a formula $A(\vec{x},\vec{y})$, a sequence of models $(M_k)$ of $\vec{y}$,
and a sequences of formulas $(I_k)$ over $\vec{y}$, we call $(I_k)$ a \emph{model
interpolation sequence} for $A$ and $(M_k)$ if for all $k$ it holds that
\begin{enumerate}
  \item $M_k$ is consistent with $\bigwedge_{i < k} I_i$;
  \item $M_k$ is inconsistent with $A$;
  \item $I_k$ is a model interpolant between $A$ and $M_k$.
\end{enumerate}
\end{definition}

\begin{definition}[Finite Convergence]
We say that \ISolverCheck{} has the \emph{finite convergence property} if it
does not allow infinite model interpolation sequences.
\end{definition}

\begin{lemma}[Termination]
\label{lm::termination}
If \ISolverCheck{} has the finite convergence property, then
\textsc{interpolate($A$, $B$)} always terminates.
\end{lemma}

\subsection{SMT Modulo Models with MCSAT}

We build a procedure for solving SMT modulo models by modifying the
satisfiability checking procedure of MCSAT. The MCSAT method for SMT
solving was introduced in~\cite{de2013model,jovanovic2013design} and
further extended in~\cite{jovanovic2017solving}. We give a brief
overview of the MCSAT terminology and mechanics, and we describe the
satisfiability procedure. We emphasize modifications to the original
MCSAT procedure that are needed for solving SMT modulo models.

The architecture of an MCSAT solver consists of a core solver, an
assignment trail, and reasoning plugins. The \emph{core solver} drives
the overall solving process, and is responsible for dispatching
notifications and handling requests from the plugins. The \emph{solver
trail} is a chronological record that tracks assignments of terms to
values. It is shared by the core solver and the reasoning
plugins. The \emph{reasoning plugins} are modules dedicated to
handling specific theory terms and constraints (e.g., clauses for
Booleans, polynomial constraints for arithmetic). A plugin reasons
about the content of the solver trail with respect to the set of
currently relevant terms. In the context of nonlinear arithmetic
problems, the reasoning plugins are the arithmetic plugin and
the Boolean plugin. The most important role of the core solver is to
perform conflict analysis when one of the reasoning plugins detects a
conflicting state.

When formulas $F_1, \ldots, F_n$ are asserted, by calling
\ISolverAssert{$F_i$}, the core solver notifies all plugins of the asserted
formulas. The plugins analyze the formulas and report
all \emph{relevant terms} back to the core. The relevant terms are the
variables and subterms of the formulas $F_i$s that need to be
consistently assigned to ensure a satisfying assignment. In nonlinear 
arithmetic, relevant  terms are all variables, arithmetic 
constraints, and non-negated Boolean terms that appear 
in the input formula (or are part of a learnt clause). Once the
relevant terms are collected, the core solver adds the assertions to
the trail. The initial trail contains then the partial assignment
$\propagation{F_i}{\btrue}{}$ and the search for a full satisfying
assignment starts from this trail.

\begin{algorithm}
\caption{
{\sc mcsat::check({\color{\emphcolor}$\vec{x} \mapsto \vec{v}$})}
}
\label{mcsat::check}
\KwData{solver trail $M$, relevant variables/terms to assign in $queue$}
\While {\True} {
	\SolverPropagate{} \;
  \uIf {a plugin detected a conflict and the conflict clause is $C$} {
  	$\langle C, final \rangle$ $\gets$ \SolverAnalyze{$M$, $C$, \color{\emphcolor}{$\vec{x}$}} \;
    \uIf {$final$} {
      \begingroup
      \color{\emphcolor}
      $I$ $\gets$ \SolverAnalyzeFinal{$M$, $C$} \;
      \endgroup
      \Return{$\langle \Unsat, {\color{\emphcolor}I} \rangle$}
    }
    \lElse {
      \SolverBacktrackWith{$M$, $C$}
    }
  }
  \Else {
    \begingroup
    \color{\emphcolor}
    \uIf {exists $x_i \in \vec{x}$ unassigned in $M$} {
      \OwnerOf{$x_i$}.\SolverDecide{$x_i$, $v_i$}
    }
    \endgroup
    \Else {
      \lIf {$queue$.\VarQueueEmpty{}} { \Return{$\langle \Sat, M\rangle$} }
      $x$ $\gets$ $queue$.\VarQueuePop{} \;
      \lIf {$x$ is unassigned} { \OwnerOf{$x$}.\SolverDecide{$x$} }
    }
	}
}
\end{algorithm}

\paragraph{Solver trail and evaluation.}

The assignment trail is the central data structure in the MCSAT
framework.  It is a generalization of the Boolean assignment trail
used in modern CDCL SAT solvers. The trail records a partial (and
potentially inconsistent) model that assigns values to relevant terms.
If the satisfiability algorithm terminates with a \Sat answer, the
full satisfying assignment can be read off the trail. At any point
during the search, the trail can be used to evaluate any relevant
compound term based on the values of its sub-terms. A term $t$ (and
$\neg t$, if Boolean) \emph{can be evaluated\/} in the trail $M$ if
$t$ itself is assigned in $M$, or if all closest relevant sub-terms of
$t$ are assigned in $M$ (and its value can therefore be computed). As
the search progresses, it is possible for some terms to \emph{be
evaluated in two different ways}, which can result in a conflict
(i.e., a term assigned different values). In order to account for this
ambiguity, we define an evaluation predicate $\evaluates{M}{t}{v}$
that returns $\True$ if the term $t$ can evaluate to the value $v$ in
trail $M$.

\paragraph{Conflicts and conflict clauses.}

One of the main responsibilities of reasoning plugins is to ensure
that the trail is consistent at any point in the search. A trail
is \emph{evaluation consistent} if no relevant term can evaluate to
two different values, as described above. A trail is \emph{unit
consistent} if every relevant term can be given a value without
making the trail evaluation inconsistent. If the trail is not
evaluation consistent or unit consistent, the trail is \emph{in
conflict}.

Trail consistency is a generalization of the consistency that CDCL SAT
solvers enforce during their search. By unit propagation, a SAT solver
ensures that, if no conflict has been detected, no clause can be
falsified by assigning a single variable (i.e., no clause evaluates to
both $\btrue$ and $\bfalse$). In the MCSAT framework, the plugins do
the same: they keep track of unit constraints and reason about the
consistency of the trail. It is the responsibility of the plugin to
report conflicts. Each conflict must be accompanied with
a \emph{valid} conflict clause that explains the
inconsistency.\footnote{By valid here we mean that the clause is a
universally true statement on its own.} A clause $C
\equiv (L_1 \vee \ldots \vee L_n)$ is a \emph{conflict clause} in a trail $M$,
if each literal $L_i$ can evaluate to $\bfalse$ in $M$, i.e. if
$\evaluates{M}{L_i}{\bfalse}$.

\begin{example}
Consider the constraint $C \equiv (x^2 + y^2 < 1)$ with the set of relevant terms $\lbrace C, x, y \rbrace$, and the following solver trails
\begin{align*}
  M_1 &= \trail{\decision{C}{\btrue}, \decision{x}{0}}\enspace, &
  M_2 &= \trail{\decision{C}{\btrue}, \decision{x}{0}, \decision{y}{0}}\enspace,\\
  M_3 &= \trail{\decision{C}{\btrue}, \decision{x}{1}}\enspace, &
  M_4 &= \trail{\decision{C}{\btrue}, \decision{x}{1}, \decision{y}{0}}\enspace.
\end{align*}
The trails $M_1$ and $M_2$ are consistent, the trail $M_3$ is unit
inconsistent (no consistent assignment for $y$ exists), and $M_4$ is
evaluation inconsistent ($C$ evaluates to both $\top$ and $\bot$).
A valid explanations for the inconsistency of $M_3$ is the conflict clause
$C_3 \equiv \neg C \vee (x < 1)$, while a valid explanation for the
inconsistency of $M_4$ is the conflict clause $C_4 \equiv \neg C \vee C$.
Although $C_4$ is a tautology, it is an acceptable conflict clause since
both literals can evaluate to $\bfalse$ (because $\evaluates{M_4}{C}{\btrue}$ and
$\evaluates{M_4}{C}{\bfalse}$).
\end{example}

\paragraph{Main procedure.}

The implementation of the satisfiability checking procedure \ISolverCheck{} is a
generalization of the search-and-resolve loop of modern SAT solvers (see, e.g.
\cite{een2003extensible,een2003temporal}). The procedure is shown in
Algorithm~\ref{mcsat::check}, where we emphasize the extensions needed for SMT
modulo models in {\color{\emphcolor}red}.
The overall procedure performs a direct search for a satisfying assignment and
terminates either by finding an assignment that extends the given partial model,
or deduces that the problem is unsatisfiable as certified by an appropriate
model interpolant.

The main elements of the procedure are unit propagation and decisions, used for
constructing the assignment, and conflict analysis for repairing the trail when
it becomes inconsistent.
The \SolverPropagate{} procedure invokes the propagation procedures provided by
the plugins. Propagation allows each plugin to add new assignments to
the top of the trail. If, during propagation, a plugin detects an inconsistency,
it reports the conflict to the core solver along with a valid conflict clause.
The \SolverDecide{$x$} procedure assigns a value of the given unassigned term
$x$. Decisions are performed only after propagation has fully saturated with no
reported conflicts, which means that the trail is unit consistent. In such a trail,
an assignment for $x$ is guaranteed to exist, but the choice of a particular
value is delegated to the plugin responsible for $x$ (e.g., the arithmetic
plugin for real-typed terms).
\begin{extension}[Decisions]
To support SMT modulo a model $\vec{x} \mapsto \vec{v}$, variables $x_i
\in \vec{x}$ of the input model are decided before any other term, and are
assigned the provided value $v_i$. The procedure that performs this decision is denoted with
\SolverDecide{$x_i$, $v_i$}.
If a decision introduces an evaluation inconsistency, the 
plugin reports the conflict with a conflict clause. %
\end{extension}
Detecting and explaining decision conflicts is straightforward: there must
exist a single constraint $C$ that can evaluate to both $\top$ and $\bot$ in
the trail. Such conflicts can always be explained with a clause of the form
$(\neg C \vee C)$.

If a conflict is reported, either during propagation or in a decision, the
procedure invokes the conflict analysis procedure \SolverAnalyze{}. This
procedure takes the reported conflict clause $C$ and finds the root cause of the
conflict. The analysis backtracks the trail, element by element, so long as $C$
is a conflict clause, while resolving any trail propagations from $C$. Once
done, the analysis returns the clause along with the flag that indicates whether
this conflict clause $C$ is empty (indicating the final conflict). If the
conflict is not final, the procedure calls \SolverBacktrackWith{} to backtrack
the trail further, if possible, and add a new assignment to the trail, ensuring
progress and fixing the conflict. The main invariant of the conflict resolution
procedure is that the \emph{conflict clause $C$ is always implied by asserted
formulas}.
\begin{extension}[Conflict Analysis]
To support SMT modulo a model $\vec{x} \mapsto \vec{v}$, the analysis
procedure \SolverAnalyze{$M$, $C$, $\vec{x}$} stops as soon as it
encounters a variable $x_i \in \vec{x}$ to resolve, and returns $\langle C,
\True\rangle$.
\end{extension}
This modification is based on the fact that the variables $x_i$ have a
fixed value given by the model. Assume that conflict analysis attempts
to undo a variable $x_i$ that is part of the provided model
$\vec{x} \mapsto \vec{v}$. This can only happen when the trail
consists of only variables from $\vec{x}$ and implications of asserted
formulas. In other words, this particular conflict cannot be resolved
unless we modify either the assertions themselves or the input
model. The clause resulting from the analysis marked as final is our
starting point for producing the model interpolant.
\begin{extension}[Final Analysis]
To support SMT modulo a model $\vec{x} \mapsto \vec{v}$, the procedure
\SolverAnalyzeFinal{$M$, $C$} resolves any remaining trail propagations in $M$
from the clause $C$ and returns the resulting clause $I$.
\end{extension}
The resolution of propagations in this final analysis is done in the
same manner as in regular conflict analysis. This means that the
resulting clause $I$ is implied by the asserted formulas. In
addition, resolving all propagations from the conflict clause
ensures that all literals of $I$ evaluate to false only because of the
assignment $\vec{x} \mapsto \vec{v}$, making $I$ an appropriate model
interpolant.

\begin{example}
\label{ex::mcsat}
Consider two formulas $F_1 \equiv b$ and $F_2 \equiv \neg b \vee (x^2 + y^2 <
2)$. When asserting these two formulas to the MCSAT solver, the Boolean and
arithmetic plugins will identify the set of terms relevant for satisfiability as
$R = \lbrace b, x, y, (x^2 + y^2 < 2) \rbrace$. Additionally, the assertions
will be added to the trail and propagated\footnote{Notation $\propagation{t}{v}{F}$ 
denotes that $t$ is assigned to $v$ due to propagation, and $F$ is the reason
of the propagation.}, 
resulting in the following initial trail
\begin{align*}
  M_0 = \trail{\propagation{b}{\btrue}{}, \propagation{F_2}{\btrue}{}, \propagation{(x^2 + y^2 < 2)}{\btrue}{F_2}}\enspace.
\end{align*}
We now apply our procedure to solve $F_1$ and $F_2$ modulo the partial model
$\lbrace x \mapsto 2 \rbrace$.

In the first iteration, no term in $R$ is unit (with only one variable
unassigned), and propagation does not infer any new facts or conflicts. The
procedure thus perform a decision on the unassigned variable $x$ of the model,
resulting in the trail
\begin{align*}
  M_1 = \trail{\propagation{b}{\btrue}{}, \propagation{F_2}{\btrue}{}, \propagation{(x^2 + y^2 < 2)}{\btrue}{F_2}, \decision{x}{2}}\enspace.
\end{align*}

In the second iteration, as $(x^2 + y^2 < 2)$ is unit in the trail $M_1$, the
arithmetic plugin examines the constraint and deduces that there is no potential
solution for $y$. This constitutes a unit inconsistency that the plugin reports,
along with the conflict clause\footnote{We use $(x > \sqrt{2})$ as a shorthand
for the extended constraint $x >_r \proot(x^2 - 2, 2, x)$.}
\begin{align*}
  C_0 \equiv \neg (x^2 + y^2 < 2) \vee \neg (x > \sqrt{2})\enspace.
\end{align*}

Conflict analysis takes clause $C_0$ and starts the resolution process.
As the top variable $x$ on the trail $M_1$ is part of the input model, the
analysis stops and reports that the clause $C_0$ is the final explanation. This
clause is valid, but not yet a model interpolant as it contains a literal with
variable $y$. We then proceed with the final analysis to remove such literals.
First, we resolve $(x^2 + y^2 < 2)$ from $C_0$ using its reason clause $F_1$,
which gives the clause $C_1 \equiv \neg b \vee \neg (x > \sqrt{2})$. Then, we
resolve $b$ from $C_1$ with an empty reason ($b$ is an assertion), resulting in
the final clause and model interpolant $I =\neg (x > \sqrt{2})$.
\end{example}

\section{Nonlinear Arithmetic}
\label{sec::nonlinear}

The general approach to interpolation presented so far is not specific
to nonlinear arithmetic.  We now tackle two practical issues that
arise in nonlinear arithmetic and we discuss the properties of our
interpolation procedure in the context of nonlinear arithmetic.
First, on nonlinear problems, as seen in Example~\ref{ex::mcsat}, the
interpolation procedure can return model interpolants that include
extended polynomial constraints. This is an artifact of the underlying
decision procedure (such as NLSAT \cite{jovanovic2012solving}) that
might use extended polynomial constraints to succinctly represent
conflict explanations. While such constraints make decision procedures
more effective, they are undesirable for interpolation: interpolants
should be described in the language of the input formulas, if
possible. Second, to use the interpolant procedure in the context of
model checking, we also need to devise a generalization procedure for
polynomial constraints.

This section uses concepts from cylindrical algebraic decomposition
(CAD).  We keep the presentation example-driven and focused on our
particular needs, and refer the reader to the existing literature for
further information
\cite{buchberger1982computer,caviness2004quantifier,basu2006algorithms}.
Cylindrical algebraic decomposition is a general approach for reasoning about
polynomials based on the following result due to Collins
\cite{collins1975quantifier}. For any set of polynomials $f_1, \ldots, f_k \in
\mathbb{Z}[x_1, \ldots, x_n]$ one can algorithmically decompose $\mathbb{R}^n$
into connected regions (called cells) such that all the polynomials
$f_j$ are sign-invariant in every cell $C_i$. This means that the
cells also maintain the truth value of any polynomial constraints over
the polynomials $f_i$, which is crucial in many reasoning techniques
for polynomial constraints.

\begin{figure*}
  \centering
  \includegraphics[width=0.49\textwidth]{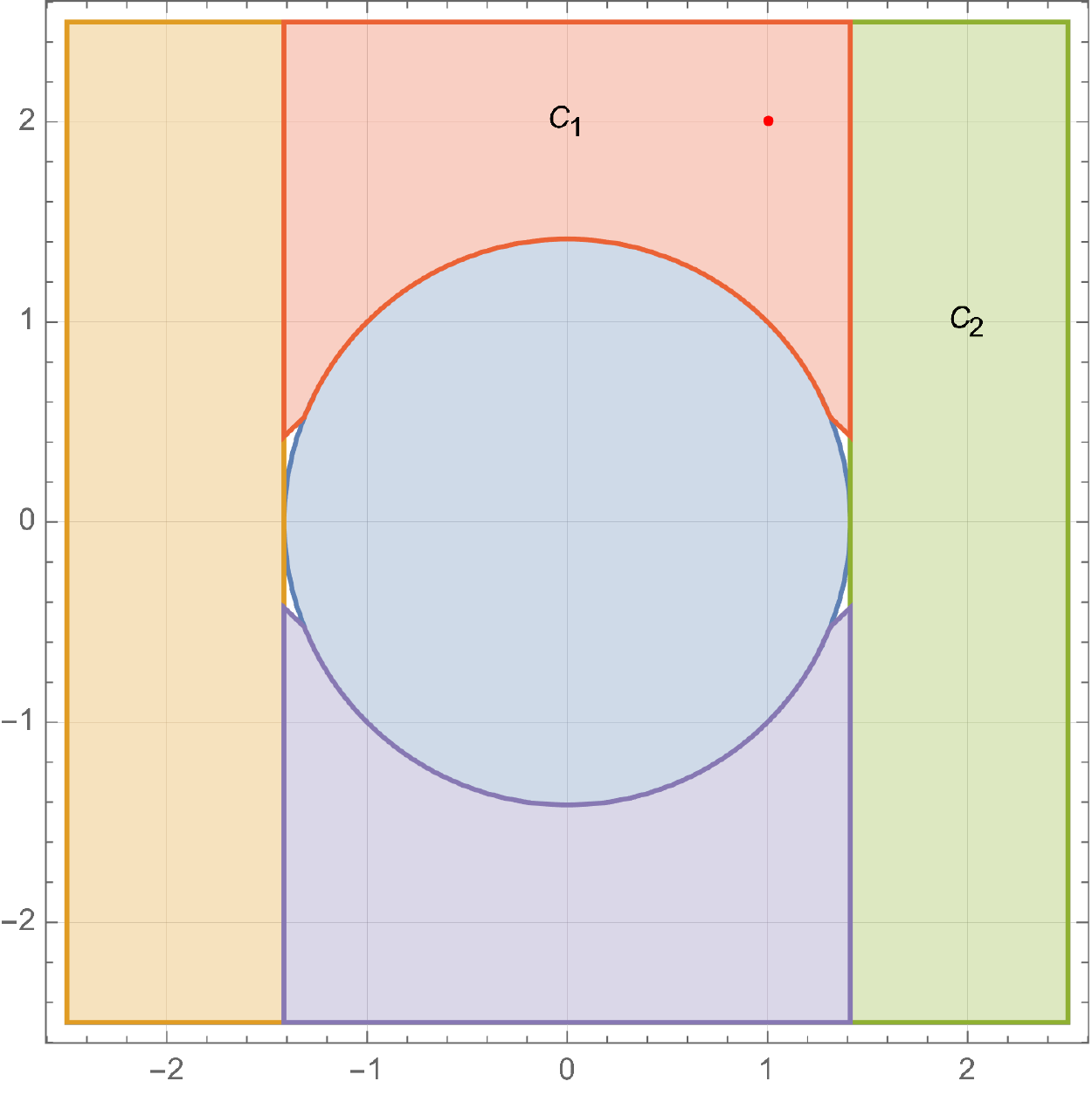}
  \includegraphics[width=0.49\textwidth]{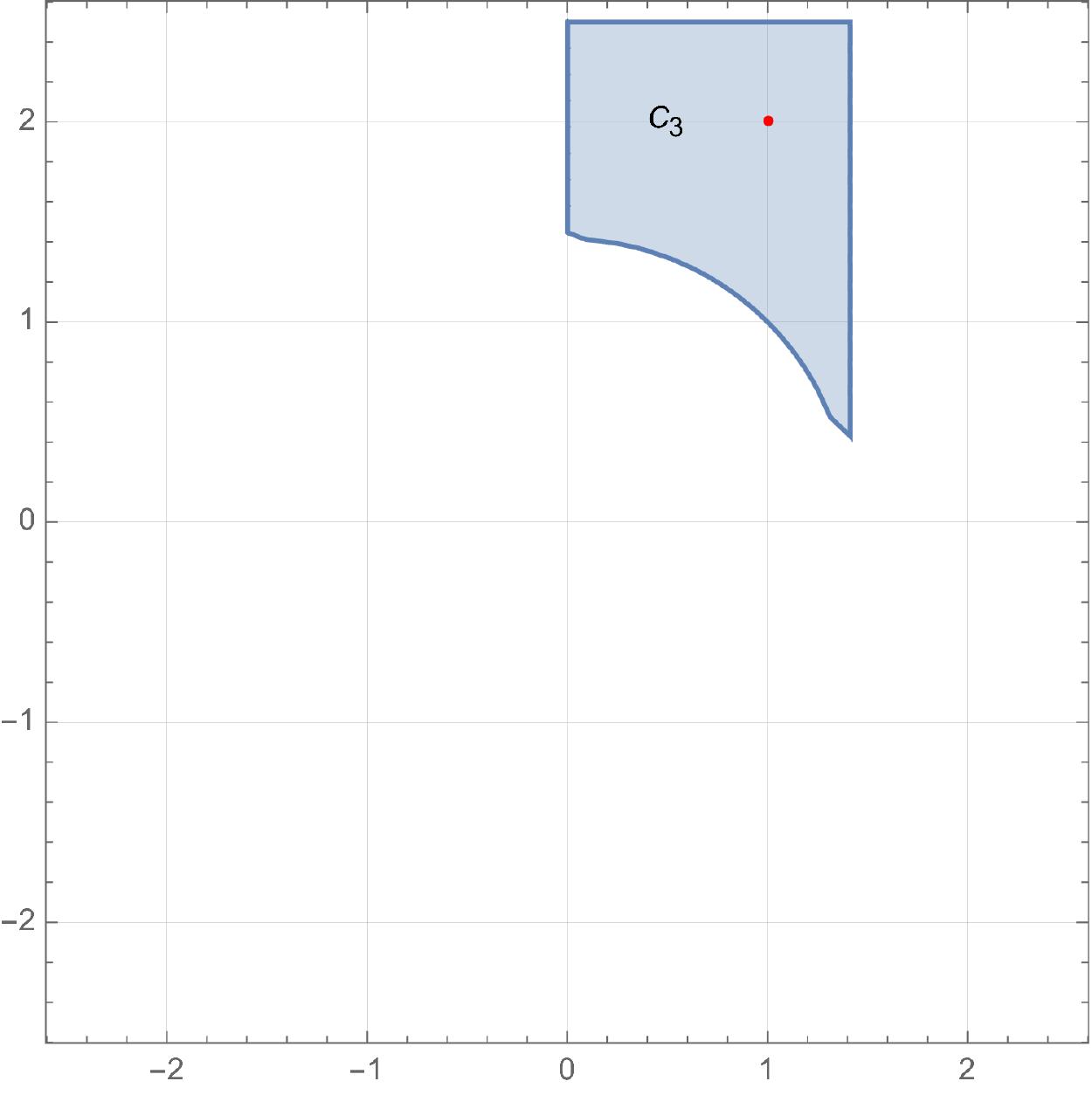}
  \caption{
  CAD of the polynomial $f = x^2 + y^2 - 2$ from Example~\ref{ex::cad} (left).
  Computed cell capturing the model $(1, 2)$ of Example~\ref{ex::cadbasic} (right).
  }
  \label{fig::decomposition}
\end{figure*}

The theory and practice of CAD is heavily dependent on the ordering of
variables involved. For this paper we always assume the CAD order to
be the same as the order of the defined polynomials (e.g., $x_1 < x_2
< \ldots < x_n$). Every CAD cell is cylindrical in nature, and can be
described by constraints where every dimension of the cell (called a
level) can be completely defined by relying only on the previous
dimensions. We illustrate this through an example.

\begin{example}
\label{ex::cad}
Consider the polynomial $f = x^2 + y^2 - 2 \in \mathbb{Z}[x, y]$. A
CAD of $f$ is depicted in Figure~\ref{fig::decomposition} (left). The
cell $C_1$ is defined by two constraints:
\begin{align*}
C^y_1 &\equiv y >_r \proot(x^2 + y^2 - 2, 2, y)\enspace, \\
C^x_1 &\equiv x >_r \proot(x^2 - 2, 1, x) \wedge x <_r \proot(x^2 - 2, 2, x)\enspace.
\end{align*}
Constraint $C^x_1$ is at the first level (it's a constraint on $x$
only), while constraint $C^y_1$ is at the second level and relates variables $x$ and $y$.
The full cell description is then $C_1 \equiv C_1^x \wedge C_1^y$. The
green cell $C_2$ can be described by $C^y_2 \equiv \top$ and
$C^x_2 \equiv x >_r \proot(x^2 - 2, 2, x)$, with the full description
$C_2 \equiv C_2^y \wedge C_2^x$.
\end{example}

Model-based decision procedures such as NLSAT rely on CAD construction but do
not construct the complete CAD decomposition. Instead, given a point in
$\mathbb{R}^n$ they can construct a single cell of a CAD in a model-driven
fashion. For more information about this approach, we refer the reader to
\cite{jovanovic2017solving,brown2015constructing}. For our purposes we abstract
the cell construction, and denote with $\cadcell{F}{M}$ the function that, given
a set of polynomials $F$, returns a description of a CAD cell of $F$ that
contains the model $M$.

Following the terminology used in CAD, we say that a non-empty connected subset of
$\mathbb{R}^k$ is a \emph{region}. A set of polynomials $\{ f_1, \ldots f_s \}
\subset \mathbb{Z}[\vec{y}, x]$, with $\vec{y} = \langle y_1, \ldots, y_n
\rangle$, is said to be \emph{delineable} in a region $S \subseteq \mathbb{R}^n$
if for every $f_i$ (and $f_j$) from the set, the following properties are
invariant for any $\vec{\alpha} \in S$:
\begin{enumerate}
\item the \emph{total number of complex roots} of $f_i(\vec{\alpha}, x)$;
\item the \emph{number of distinct complex roots} of $f_i(\vec{\alpha}, x)$;
\item the \emph{number of common complex roots} of $f_i(\vec{\alpha}, x)$ and $f_j(\vec{\alpha}, x)$.
\end{enumerate}
Delineability has important consequences on the number and arrangement of real
roots of polynomials $f_i$. As explained by the following theorem, if a set of
polynomials $F$ is delineable on a region $S$, then the number of real roots of
the polynomials does not change on $S$. Moreover, these roots maintain their
relative order on the whole of $S$.

\begin{theorem}[Corollary 8.6.5 of \cite{mishra1993algorithmic}]
\label{th::delineability}
Let $F$ be a set of polynomials in $\mathbb{Z}[\vec{y}, x]$, delineable in a
region $S \subset \mathbb{R}^n$. Then, the real roots of $F$ vary continuously
over $S$, while maintaining their order.
\end{theorem}
For a polynomial $f \in \mathbb{Z}[\vec{x}]$ and model $M = \lbrace \vec{x}
\mapsto \vec{v} \rbrace$, we denote with $\sgncstr(f, M)$ the polynomial
constraint that matches the sign of $f$ in $M$, i.e.
\begin{align*}
\sgncstr(f, M) = \begin{cases}
f < 0 & \text{ if } \sgn(f(\vec{v})) < 0 \\
f > 0 & \text{ if } \sgn(f(\vec{v})) > 0 \\
f = 0 & \text{ if } \sgn(f(\vec{v})) = 0 \\
\end{cases}
\end{align*}
As described above, a CAD cell can be succinctly described by relying on
extended polynomial constraints. We now show that the description of the cell
can be reduced to basic polynomial constraints.

\begin{lemma}
\label{lm::simplify}
Let $f_i \in \mathbb{Z}[y_1, \ldots, y_n, x]$ be two polynomials of degrees
$m_i$, and $F_i \equiv x \; \triangledown_r \; \proot(f_i, k_i, x)$ be extended
polynomial constraints of a cell description. Let $S$ be a region of
$\mathbb{R}^n$ where $\lbrace f_1, f_2 \rbrace$ are delineable and let $M =
\lbrace \vec{y} \mapsto \vec{v}, x \mapsto \alpha \rbrace$ be a model such that
$\vec{v} \in S$. Then, for all $\vec{y} \in S$ it holds that
\begin{align*}
  \bigwedge_{i = 0}^{m_1-1} \sgncstr(f_1^{(i)}, M) \wedge \bigwedge_{i = 0}^{m_2-1} \sgncstr(f_2^{(i)}, M) \Rightarrow F_1 \wedge F_2 \enspace.
\end{align*}
\end{lemma}

The proof of this lemma is relatively straightforward. The CAD cell description
for level $x$ represents an entry in the sign table of $f_1$ and $f_2$ (with no
roots in between). A part of this sign table entry that contains $M$ can be
described with the signs of all the derivatives of $f_1$ and $f_2$ as long as we
can guarantee that neither the arrangement nor the number of roots $f_1$ and
$f_2$ change. But, this is guaranteed by $f_1$ and $f_2$ being delineable on
$S$, so the lemma holds.

As a corollary to this lemma, in the context of CAD cell construction
around a model $M$, we can replace any extended constraints describing
a cell $C$ with basic constraints stating that the signs of the
polynomial derivatives are the same as in $M$. This results in a valid
CAD subcell $C' \subseteq C$ for the same polynomials, that still
contains the model $M$. We denote the function that constructs a
basic CAD cell description of a set of polynomials $F$ capturing the
model $M$ with $\cadcellbasic{F}{M}$.

\begin{example}
\label{ex::cadbasic}
Based on Example~\ref{ex::cad}, we can construct a cell around the
model $M = \lbrace x \mapsto 1, y \mapsto 2\rbrace$. Function
$\cadcellbasic{F}{M}$ will return the constraints
\begin{align*}
C^y_3 &\equiv (x^2 + y^2 > 2) \wedge (y > 0)\enspace, \\
C^x_3 &\equiv (x^2 < 2) \wedge (x > 0)\enspace.
\end{align*}
The full cell description is then $C_3 \equiv C_3^x \wedge C_3^y$. Note that
this cell is smaller than the cell $C_1$ from Example~\ref{ex::cad}. This
reduction in size is generally undesirable, but it is a price to pay for having
the description in a simpler language.
\end{example}

\paragraph{Interpolation without extended constraints.}

We now show how the cell construction described above can be used to
remove extended polynomial constraints from a model interpolant.
Assume a clausal model interpolant
\begin{align*}
  I = (L_1 \vee \ldots \vee L_i \vee \ldots \vee L_N)
\end{align*}
that is implied by formula $A$ and refutes a model $M = \lbrace \vec{x} \mapsto
\vec{v} \rbrace$, i.e., all literals of $I$ evaluate to $\bot$ in $M$. Assume
also that some literal $L_i$ contains an extended polynomial
constraint $x_n \; \triangledown_r \; \proot(f, k, x_n)$, with $f \in
\mathbb{Z}[\vec{x}]$. We aim to replace the extended literal $L_i$ with literals
over basic polynomial constraints. To do so, we need to find literals $L^1_i,
\ldots, L^m_i$ such that $L_i \Rightarrow (L^1_i \vee \ldots \vee L^m_i)$ and
all literals $L^j_i$ evaluate to $\bfalse$ in $M$. Then, the clause
\begin{align*}
I' = (L_1 \vee \ldots \vee L^1_i \vee \ldots L^m_i \vee \ldots \vee L_N)
\end{align*}
will also be a model interpolant implied by $A$ that refutes the model $M$.

We can construct the literals $L^j_i$ using single cell construction as follows.
We create a description of the CAD cell of the polynomial $f$ from $L_i$ that
captures the model $M$. Let $\cadcellbasic{\lbrace f \rbrace}{M} = D_1 \wedge
\ldots \wedge D_m$ be this description. Since the cell fully captures the
behavior of $f$ around $M$, we know that $D_1 \wedge \ldots
\wedge D_m \Rightarrow \neg L_i$ and all literals $D_j$ evaluate to $\btrue$.
Therefore, we can use the cell description to eliminate the extended literal
$L_i$, obtaining the clause
\begin{align*}
I' = (L_1 \vee \ldots \vee \neg D_i \vee \ldots \neg D_m \vee \ldots \vee L_n)
\end{align*}

By continuing this process, we can replace all extended literals from a model
interpolant, to obtain a model interpolant in the basic language of polynomial
constraints.

\begin{example}
Consider the model interpolant $I = \neg (x >_r \proot(x^2 - 2, 2, x)$ from
Example~\ref{ex::mcsat} that refutes the model $M = \lbrace x \mapsto 2
\rbrace$. To express $I$ in terms of basic polynomials constraints we first
construct a regular CAD cell of $f = x^2 - 2$ around $M$. In this case this cell
is simply $x >_r \proot(x^2 - 2, 2, x)$. Then, we use Lemma~\ref{lm::simplify} to
construct a basic CAD cell description as $(x^2 > 2) \wedge (x > 0)$. Finally,
the simplified interpolant is $I' = \neg (x^2 > 2) \vee \neg (x > 0)$.
\end{example}

\paragraph{Termination.}

With the description of the interpolation procedure complete, we discuss the
termination of the procedure. To do so, we fix the formula $A(\vec{x}, \vec{y})$
of Definition~\ref{def::seq} and we assume a fixed order of variables that ensures
$y_i < x_i$. Since the MCSAT decision procedure on which we rely is based on CAD,
we can put a bound on the set of literals that can ever appear in a model
interpolant from the formula $A$ to an arbitrary model $M$. Let $P_A$ be the set
of polynomials appearing in $A$, and let $P = \mathsf{P}(P_A)$ denote the
closure of the set $P_A$ under the CAD projection operator used by the decision
procedure. Finally, let $P'$ be the closure of $P$ under derivatives. The set of
polynomial constraints that can appear in the interpolant $I$ is limited to
basic polynomial constraints over polynomials in $P'$.
This means that the procedure {\sc mcsat::check()} can only generate a
finite number of model interpolants and therefore has the finite
convergence property.

\begin{lemma}
Assuming a fixed variable order, the {\sc mcsat::check()} procedure has the
finite convergence property for nonlinear arithmetic formulas.
\end{lemma}
Together with Lemma~\ref{lm::termination}, this lemma implies that our
interpolation procedure for the theory of nonlinear arithmetic
terminates.

\paragraph{Model generalization.}

We now proceed to show how the CAD cell construction can be used in a natural
way to provide model-driven generalization. As in Definition~\ref{def::gen},
assume a formula $F(\vec{x}, \vec{y})$ such that $F$ is true in a model $M$. Our
aim is to construct a formula $G(\vec{x})$ that generalizes the model $M$ and
still guarantees a solution to $F$.

Following the approach of \cite{dutertre2015solving}, we do this in
two steps.  First, we construct an implicant $B$ of $F$ based on the
model $M$. Then, we eliminate the variables $\vec{y}$ from $B$, again
relying on the model $M$. The implicant $B$ is a conjunction of
literals that implies $F$ and such that $B$ is true in M. The
implicant can be computed by a top-down traversal of the
formula $F$ while using the model $M$ to evaluate the formula nodes
(see, e.g., \cite{dutertre2015solving} for a detailed description). To
find a formula $G$ such that $G \Rightarrow \exists \vec{y} \;.\; B$,
we use CAD cell construction as follows. Let
$P \subseteq \mathbb{Z}[\vec{x}, \vec{y}]$ be the set of all
polynomials appearing in $B$, and let the cell description of $P$
around $M$ be
\begin{align*}
  \cadcellbasic{P}{M} = D_{\vec{x}} \wedge D_{\vec{y}}\enspace.
\end{align*}
Here, $D_{\vec{x}}$ denotes the description of cell levels of
variables $\vec{x}$, while $D_{\vec{y}}$ denotes the description of
cell levels of variables $\vec{y}$. Because of the cylindrical nature
of CAD cells, and the order on variables $y_i$ and $x_i$, we are
guaranteed that every solution of $D_{\vec{x}}$ can be extended to a
solution of $D_{\vec{y}}$. Therefore we set the final generalization
$G(x) \equiv D_{\vec{x}}$

\begin{example}[Generalization]
Consider the formula $F \equiv (x^2 + y^2 < 2)$ and the model $M = \lbrace x
\mapsto 1, y \mapsto 2 \rbrace$ that satisfies $F$, and let us compute a
generalization $G(x)$ of $M$. First, we compute a CAD cell of $f = x^2 + y^2 -
2$ as shown in Example~\ref{ex::cadbasic}. Then we drop the description of cell
level $y$, to obtain the model generalization $G(x) \equiv (x^2 < 2) \wedge (x >
0)$.
\end{example}

\section{Evaluation}
\label{sec::evaluation}

To the best of our knowledge, there is no clear metric for evaluating how good
an interpolant is, or for comparing different interpolants. In this section, we
first show two examples to illustrate the procedure and its applications. Then,
we evaluate the effectiveness of our interpolation procedure on practical
problems that arise from model-checking applications. To this end, we integrate
the procedure into a model checker and evaluate whether the procedure is
efficient, and can produce abstractions that help the model checker synthesize
invariants and discover counter-examples.

We have implemented the reasoning procedures (solving modulo partial models and
interpolation procedure) by extending the existing \mcsat implementation of the
\yices SMT solver \cite{dutertre2014yices}. We used the \libpoly library
\cite{jovanovic2017libpoly} for computing the model generalization and
simplification of algebraic cells. Since \yices is integrated into the \sally
model checker \cite{jovanovic2016property}, we rely on the \pdkind method
\cite{jovanovic2016property} as the model checking engine (the user of
interpolation) in our evaluation.

\begin{figure*}
  \centering
  \includegraphics[width=0.45\textwidth]{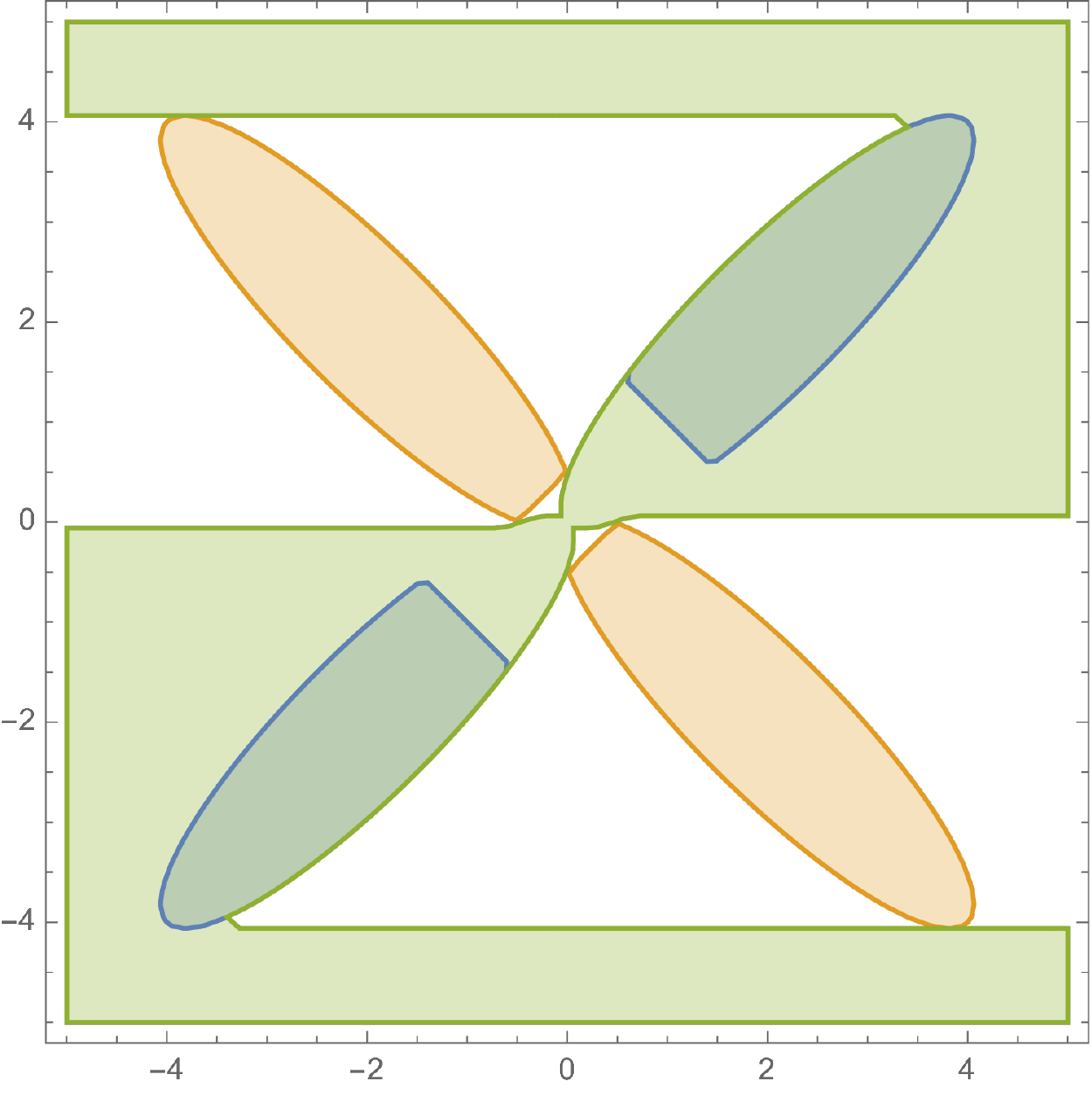}
  \;\;\;\;
  \includegraphics[width=0.45\textwidth]{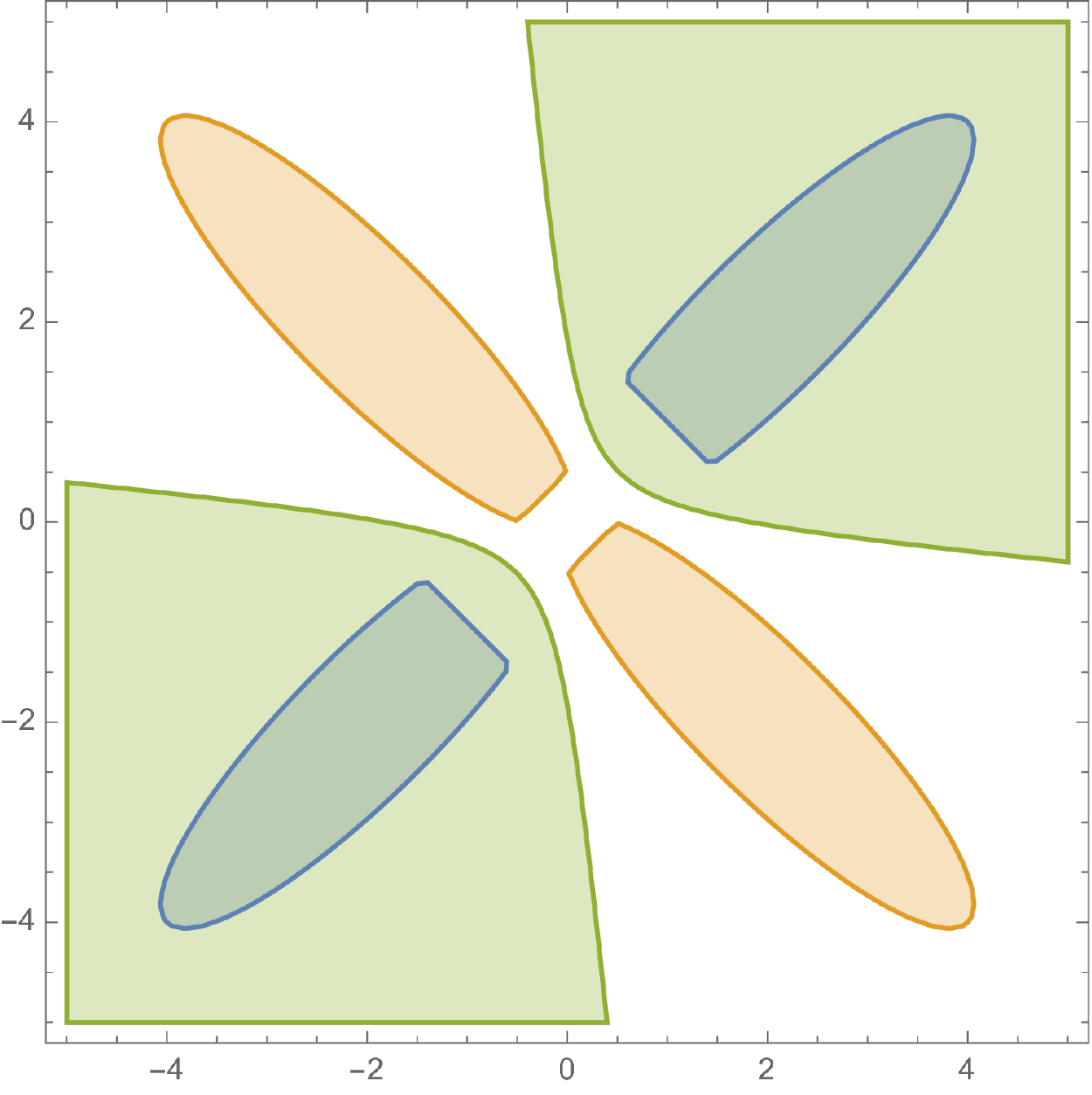}
  \caption{
  Illustration of interpolants from Example~\ref{ex::gan2020}. In blue and
  orange are the feasible space of the formulas $A$ and $B$ (projected on $x$
  and $y$). In green is the feasible space of the interpolant produced by our
  method (on the left) and the interpolant produced by \cite{gan2020nonlinear}
  (on the right).
  }
  \label{fig::gan2020}
\end{figure*}

\begin{example}
\label{ex::gan2020}
We compare the style of interpolants generated by our new procedure with the
ones generated by numerical approaches such as \cite{gan2020nonlinear}.
Example~4 from \cite{gan2020nonlinear} considers two formulas of the form
\begin{align*}
  A(x, y, a_1, a_2, b_1, b_2) \equiv (f_1 \geq 0 \wedge f_2 \geq 0) \vee (f_3 \geq 0 \wedge f_4 \geq 0)\enspace, \\
  B(x, y, c_1, c_2, d_1, d_2) \equiv (g_1 \geq 0 \wedge g_2 \geq 0) \vee (g_3 \geq 0 \wedge g_4 \geq 0)\enspace.
\end{align*}
The polynomials $f_i$ and $g_i$ involved in $A$ and $B$ are of degree
2. The right-hand side of Figure~\ref{fig::gan2020} shows the
interpolant $I_1$ found by the approach
in \cite{gan2020nonlinear}. This interpolant is of the form $h(x, y) >
0$, where $h$ is a polynomial degree two computed using semidefinite
programming. Our approach, on the other hand, produces the interpolant
$I_2$ shown on the left-hand side of Figure~\ref{fig::gan2020}. This
interpolant consists of 12 clauses, each containing 6--8 polynomial
constraints over 16 different polynomials (8 linear, 8 of degree
2). The interpolant $I_2$ is ultimately produced from fragments of a
CAD so its edges touch upon the critical points of the shape they were
produce from (formula $A$). Interpolant $I_1$, on the other hand, has
a simple form dictated by the method \cite{gan2020nonlinear}. Which
form is ultimately more useful depends on a particular application.
\end{example}

\begin{example}[Cauchy-Schwartz]
As described in Example~\ref{ex::cauchy}, we can model the computation of
Cauchy-Schwarz inequality as a transition system $\system_{cs}$. Then we can
prove the inequality correct if we can prove that the property $P_{cs}$ is valid
in $\system_{cs}$. The \pdkind model checking engine with the new interpolation
procedure proves the property valid in 1s.
\end{example}

\paragraph{Benchmarks.}

We run the evaluation on an existing set of nonlinear model-checking
problems used by Cimatti, et~al.~\cite{cimatti2017invariant}. This set
consists of 114 benchmarks from various sources: handcrafted
benchmarks, hybrid system verification, \nuxmv benchmarks, C
floating-point verification, and verification of Simulink models.  The
benchmark problems all contain transition systems with nonlinear
behavior. For each problem, the goal is to prove or disprove a single
invariant. We refer the reader to \cite{cimatti2017invariant} for a
more detailed description.

\paragraph{Evaluation.}

Cimatti, et~al.~\cite{cimatti2017invariant} present an abstraction approach
based on incrementally more precise linear approximations of nonlinear
polynomials. They show that this approach, implemented in the \icccnra\ tool, is
superior to other tools (such as, \isat \cite{mahdi2016advancing} and
\nuxmv \cite{cavada2014nuxmv} with upfront linear abstraction). Since our goal
is to show the effectiveness of our interpolation procedure, rather than compare
to many model checking engines, we keep the evaluation simple and only compare
to \icccnra. In addition, we include the k-induction engine \kind of \sally  in the
comparison  to illustrate  the importance of invariant inference and
counter-example generation.\footnote{\kind performs $k$-induction checks 
for increasing values of $k$ and stops if either the property is shown $k$-inductive, 
or a counter-example is found.}

\begin{figure*}
  \centering
  \includegraphics[width=0.49\textwidth]{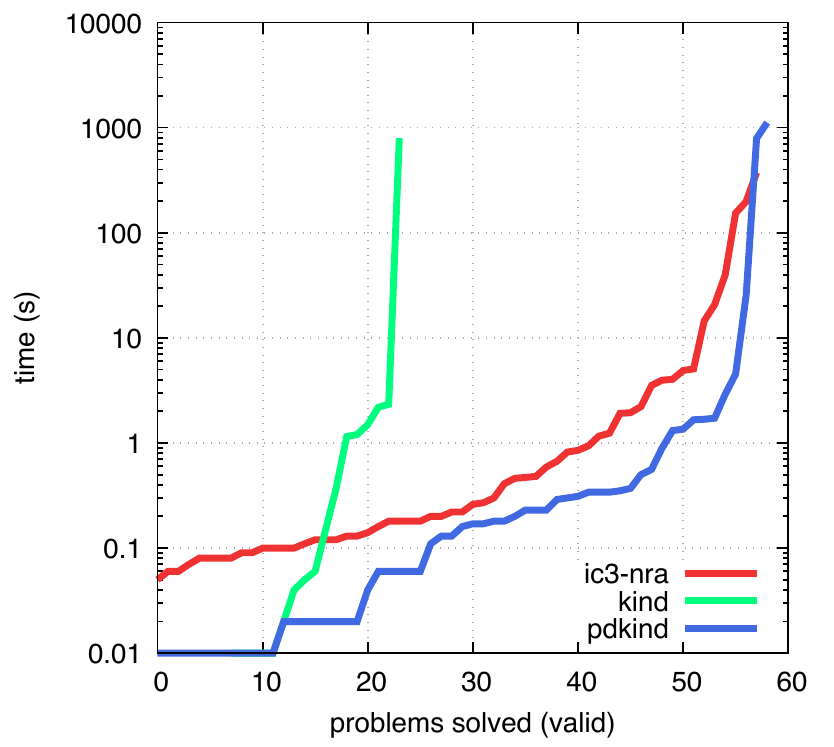}
  \includegraphics[width=0.49\textwidth]{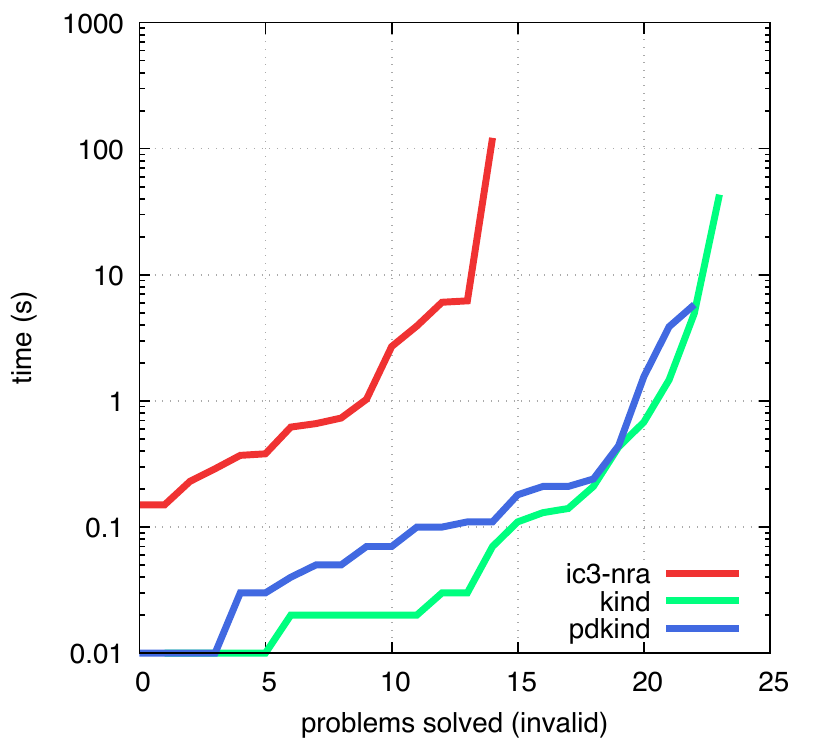}
  \caption{
  Cactus Plots Comparing the Performance of \icccnra, \kind, and \pdkind. The $x$
  axis is the number of problems solved (valid on the left, invalid on the right) and the
  $y$ axis is the time needed to solve the problem (log scale).
  }
  \label{fig::cactus}
\end{figure*}

\begin{figure*}
  \centering
  \includegraphics[width=0.45\textwidth]{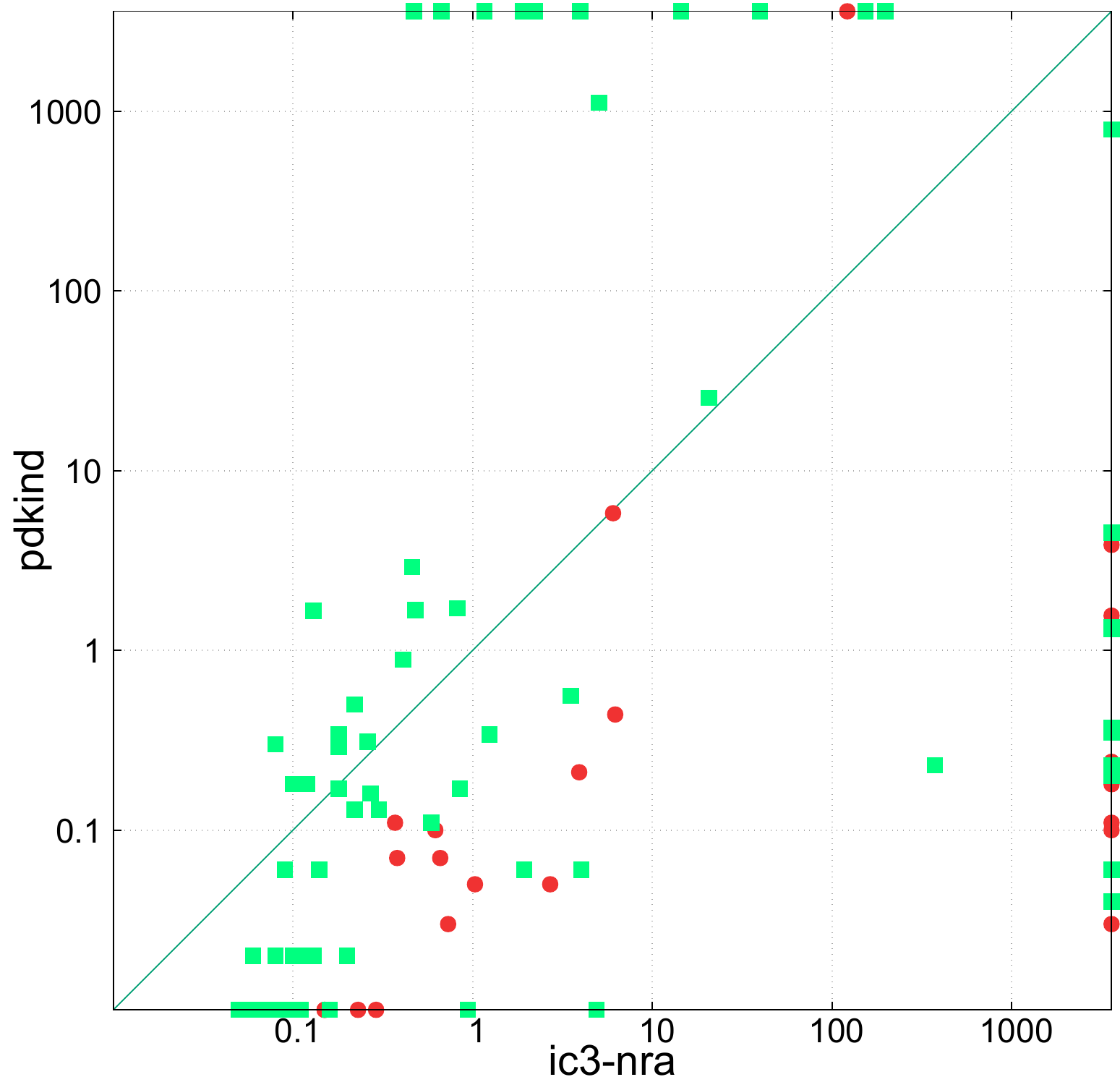}
  \;\;\;\;
  \includegraphics[width=0.45\textwidth]{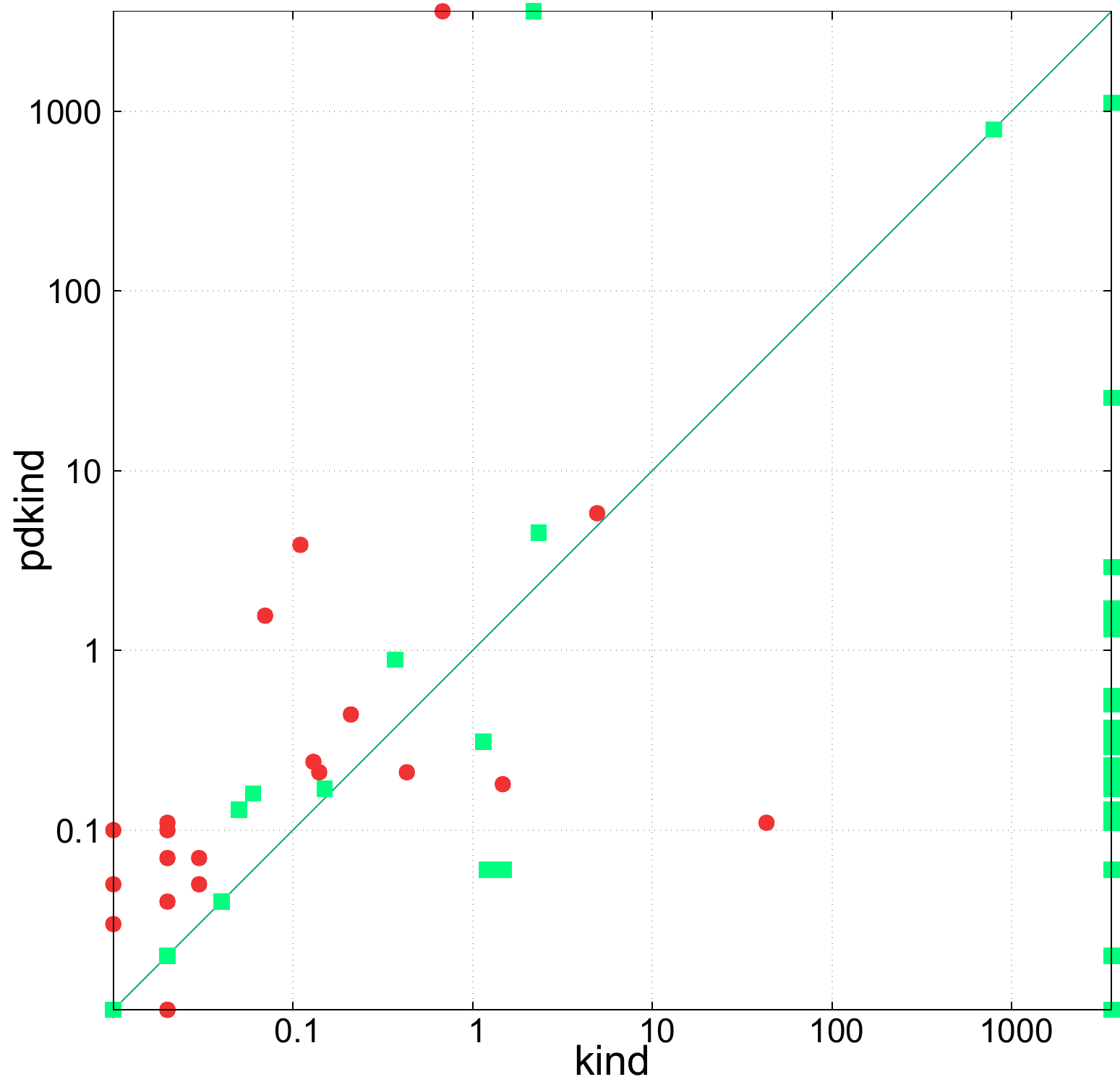}
  \caption{
  Scatter Plots Comparing the Performance of \icccnra and \kind with \pdkind.
  Green squares represent problems that are valid. Red dots represent problems
  that are invalid. Each axis represents the time it took the tool to solve
  the problem (log scale).
  }
  \label{fig::scatter}
\end{figure*}

We ran the tools on the benchmark set with a 1h CPU timeout per
problem. The results are shown in Table~\ref{tbl::results} and on the
cactus plot in Figure~\ref{fig::cactus}. A scatter plot comparison of
\pdkind against \icccnra and \kind is shown in Figure~\ref{fig::scatter}.

\begin{figure*}[t]
\caption{
Evaluation Results. For each tool, we report the number of solved
problems, how many of the solved problems were valid and invalid, and the total
time used to solve them. The rows correspond to different problem classes,
and the bottom row reports the overall results for all 114 benchmarks.
}
\label{tbl::results}
\centering

\setlength\tabcolsep{5pt}.
\begin{tabular}{| l | r r r | r r r | r r r |}
\cline{2-10}
\multicolumn{1}{c|}{} & \multicolumn{3}{c|}{\icccnra} & \multicolumn{3}{c|}{\kind} & \multicolumn{3}{c|}{\pdkind}\\
\cline{2-10}
\multicolumn{1}{l}{\tiny\textsf{problem set}}
& \multicolumn{1}{r}{\tiny\textsf{solved}} & \multicolumn{1}{r}{\tiny\textsf{valid/invalid}} & \multicolumn{1}{r}{\tiny\textsf{time (s)}}
& \multicolumn{1}{r}{\tiny\textsf{solved}} & \multicolumn{1}{r}{\tiny\textsf{valid/invalid}} & \multicolumn{1}{r}{\tiny\textsf{time (s)}}
& \multicolumn{1}{r}{\tiny\textsf{solved}} & \multicolumn{1}{r}{\tiny\textsf{valid/invalid}} & \multicolumn{1}{r}{\tiny\textsf{time (s)}}
\\
\hline
\textsf{handcrafted} (14)
 &   10 & 9/1 &   381 &    3 & 2/1 &     0 &   \textbf{14} & 13/1 &     4 \\
\rowcolor[gray]{.95}
\textsf{hycomp} (7)
 &    2 & 2/0 &    15 &    4 & 1/3 &   796 &    \textbf{4} & 2/2 &   792 \\
\textsf{hyst} (65)
 &   \textbf{39} & 32/7 &   404 &   25 & 13/12 &    50 &   38 & 26/12 &    42 \\
\rowcolor[gray]{.95}
\textsf{isat3} (1)
 &    0 & 0/0 &     0 &    0 & 0/0 &     0 &    0 & 0/0 &     0 \\
\textsf{isat3-cfg} (10)
 &    8 & 6/2 &    14 &    9 & 6/3 &     9 &   \textbf{10} & 7/3 &     8 \\
\rowcolor[gray]{.95}
\textsf{nuxmv} (2)
 &    \textbf{2} & 2/0 &   158 &    0 & 0/0 &     0 &    1 & 1/0 &  1118 \\
\textsf{sas13} (13)
 &   10 & 5/5 &    13 &    5 & 0/5 &     0 &   \textbf{13} & 8/5 &     7 \\
\rowcolor[gray]{.95}
\textsf{tcm} (2)
 &    2 & 2/0 &     1 &    \textbf{2} & 2/0 &     0 &    \textbf{2} & 2/0 &     0 \\
\hline
\multicolumn{1}{c|}{}
 &   73 &   58/15 &  986 &   48 &   24/24 &  855 &   \textbf{82} &   59/23 & 1971\\
\cline{2-10}
\end{tabular}

\end{figure*}

As can be seen from Table~\ref{tbl::results}, the results are positive. The
\pdkind engine with the new interpolation method can prove more properties
and find more counter-examples than the state-of-the-art \icccnra.

Out of 59 properties that \pdkind shows correct, 36 cannot be proved
by \kind.  This means that these properties are likely not
$k$-inductive and that the interpolants produced by our procedure are
valuable abstractions in invariant inference. Similarly, \icccnra
proves 37 properties that are not $k$-inductive.  As can be seen from
the scatter plot in Figure~\ref{fig::scatter}, there are properties
that \pdkind can prove than \icccnra cannot, and vice versa (11 and
10, respectively). This is to be expected from a difficult domain, but
it also means that the interpolation and the abstraction approach (or
other methods) can be used to complement each other.

As for the invalid properties, since our interpolation method (and thus \pdkind)
is based on complete and precise reasoning, while \icccnra relies on
abstraction, it is to be expected that \pdkind can prove more properties
invalid. Furthermore, the comparison with \kind in Figure~\ref{fig::scatter}
shows that \pdkind finds all but one counter-examples that \kind does in a
similar amount of time. We see this as a confirmation that the
interpolation and generalization methods are effective, i.e., they do not impede
the search for counter-examples.

\subsection{Related work}

There is ample literature on interpolation for different fragments of
nonlinear arithmetic. Existing methods can roughly be classified into two
categories: approaches based on interval reasoning, and approaches
based on semidefinite programming.
Interval reasoning techniques (e.g.,
\cite{kupferschmid2011craig,gao2016interpolants,mahdi2016advancing}) construct a
proof of unsatisfiability through interval slicing and
propagation. From such a proof, interpolants can be built using
proof-based interpolation techniques.  While incomplete,
interval-based techniques can be very effective on problems that are
hard for complete techniques. Moreover they can support more
polynomial functions (e.g., elementary functions, ODEs). Our procedure
is complete, but it is limited to the theories supported by MCSAT.
The approaches based on semidefinite programming
\cite{dai2013generating,gan2016interpolant,gan2020nonlinear} generally approach
the interpolation problem by restricting both the fragment of
arithmetic (e.g., bounded constraints, same set of variables,
quadratic constraints) and the shape of the interpolant (a single
polynomial constraint) so that the interpolant itself can be
represented as a semidefinite optimization problem. When they apply,
these procedures are also very effective but they suffer from
numerical imprecision, requiring special care to account for these
errors and making them difficult to use in formal verification. In
contrast, out procedure applies to nonlinear arithmetic as a whole. It
relies on symbolic techniques, which are not subject to numerical
errors. It is precise and complete, and it produces clausal interpolants.

The core ideas beyond our model-based interpolation approach were
presented at the Boolean level as SAT solving with assumptions
\cite{een2003temporal}. Closest to our work is the work of Schindler and Jovanovi\'{c} 
\cite{schindler2018selfless} where a similar model-based approach to
interpolation is applied to conjunctions of linear arithmetic
constraints based on conflict resolution. Our work is more general as
it applies to formulas other than conjunctions, and it is applicable to
a wider range of theories.

\section{Conclusion and Future Work}
\label{sec::conclusion}

We have presented a general approach for interpolation in SMT. This novel
approach relies on a mode of interaction with the SMT solver that can check a
formula for satisfiability modulo a partial model and, if the formula is
unsatisfiable, can return a model interpolant that refutes the model. This
allows us to develop a first complete interpolation procedure for nonlinear
arithmetic. We have implemented the new procedure in the \yices SMT solver and
evaluated the interpolation procedure on model-checking problems. The new
procedure seems to be effective in practice and opens new possibilities in the
verification of systems that contain nonlinear behavior. Additionally, we show
interesting examples of how the procedure can be used in automating induction
proofs in mathematics.

The interpolation procedure that we presented can support other theories
available in MCSAT (e.g., uninterpreted functions \cite{jovanovic2013design},
bit-vectors \cite{graham2020solving}, nonlinear integer arithmetic
\cite{jovanovic2017solving}). We plan to explore interpolation in these theories
in more detail, and in the contexts where interpolation can be beneficial (e.g.,
model checking, quantified reasoning, termination, and proof generation).

\bibliographystyle{abbrv}
\bibliography{main}

\end{document}